\documentstyle[prl,aps]{revtex} \topskip 2.0truecm
\def\be{\begin{equation}}
\def\ee{\end{equation}}
\def\bea{\begin{eqnarray}}
\def\eea{\end{eqnarray}}
\def\ba{\begin{array}}
\def\ea{\end{array}}

\begin{document}
\title{Chiral constituent quark model and the coupling strength of $\eta'$}
\author{Harleen Dahiya$^a$, Manmohan Gupta$^a$ and J.M.S. Rana$^{b}$}
\address {$^a$Department of Physics (Centre of Advanced Study in
Physics), Panjab University, Chandigarh-160 014, India.
\\$^b$
Department of Physics, H.N.B. Garhwal University, SRT Campus,
Badshahithaul(Tehri-Garhwal), India.}
\date{\today}
\maketitle

\begin{abstract}

Using the latest data pertaining to $\bar u-\bar d$ asymmetry and
the spin polarization functions, detailed implications of the
possible values of the coupling strength of the singlet Goldstone
boson $\eta'$ have been investigated in the $\chi$CQM with
configuration mixing. Using $\Delta u$, $\Delta_3$, $\bar u-\bar
d$ and $\bar u/\bar d$, the possible ranges of the coupling
parameters $a$, $\alpha^2 a$, $\beta^2 a$ and $\zeta^2 a$,
representing respectively the probabilities of fluctuations to
pions, $K$, $\eta$ and $\eta^{'}$, are shown to be $0.10 \lesssim
a \lesssim 0.14$, $0.2\lesssim \alpha \lesssim 0.5$, $0.2\lesssim
\beta \lesssim 0.7$ and $0.10\lesssim |\zeta| \lesssim 0.70$. To
further constrain the coupling strength of $\eta'$, detailed fits
have been obtained for spin polarization functions, quark
distribution functions and baryon octet magnetic moments
corresponding to the following sets of parameters: $a=0.1$,
$\alpha=0.4$, $\beta=0.7$, $|\zeta|=0.65$ (Case I); $a=0.1$,
$\alpha=0.4$, $\beta=0.6$, $|\zeta|=0.70$ (Case II); $a=0.14$,
$\alpha=0.4$, $\beta=0.2$, $\zeta=0$ (Case III) and $a=0.13$,
$\alpha=\beta=0.45$, $|\zeta|=0.10$ (Case IV). Case I represents
the calculations where $a$ is fixed to be 0.1, in accordance with
earlier calculations, whereas other parameters are treated free
and the Case IV represents our best fit. The fits clearly
establish that a small non-zero value of the coupling of $\eta'$
is preferred over the higher values of $\eta'$ as well as when
$\zeta=0$, the latter implying the absence of $\eta'$ from the
dynamics of $\chi$CQM. Our best fit achieves an overall excellent
fit to the data, in particular the fit for $\Delta u$, $\Delta d$,
$\Delta_8$ as well as the magnetic moments $\mu_{n}$,
$\mu_{\Sigma^-}$, $\mu_{\Sigma^+}$ and $\mu_{\Xi^-}$ is almost
perfect, the $\mu_{\Xi^-}$ being a difficult case for most of the
similar calculations.

\end{abstract}

The chiral constituent quark model ($\chi$CQM), as formulated by
Manohar and Georgi \cite{{manohar}}, has recently got good deal of
attention \cite{{eichten},{cheng},{song},{johan}} as it is
successful in not only explaining the ``proton spin crisis''
\cite{emc} but is also able to account for the $\bar u-\bar d$
asymmetry {\cite{{nmc},{e866},{GSR}}}, existence of significant
strange quark content $\bar s$ in the nucleon,  various quark
flavor contributions to the proton spin {\cite{eichten}}, baryon
magnetic moments {\cite{{eichten},{cheng}}} and hyperon
$\beta-$decay parameters etc.. Recently, it has been shown that
configuration mixing generated by spin-spin forces
\cite{{dgg},{Isgur},{yaouanc}}, known to be compatible with the
$\chi$CQM \cite{riska,chengspin,prl}, improves the predictions of
$\chi$CQM regarding the quark distribution functions and the spin
polarization functions \cite{hd}. Further, $\chi$CQM with
configuration mixing (henceforth to be referred as
$\chi$CQM$_{{\rm config}}$) when coupled with the quark sea
polarization and orbital angular
 momentum (Cheng-Li mechanism \cite{{cheng1}}) as well as
``confinement effects'' is able to give an excellent fit \cite{hd}
to the baryon magnetic moments and a perfect fit for the violation
of Coleman Glashow sum rule \cite{cg}.

The key to understand the ``proton spin problem'', in the
$\chi$CQM formalism \cite{cheng}, is the fluctuation process
\be
  q^{\pm} \rightarrow {\rm GB}
  + q^{' \mp} \rightarrow  (q \bar q^{'})
  +q^{'\mp}\,, \label{basic}
\ee
 where GB represents the Goldstone boson and $q \bar q^{'}  +q^{'}$
 constitute the ``quark sea'' \cite{cheng,song,johan,hd}.
The effective Lagrangian describing interaction between quarks and
a nonet of GBs, consisting of octet and a singlet, can be
expressed as
\be
{\cal L}= g_8 {\bf \bar q}\Phi {\bf q} + g_1{\bf \bar
q}\frac{\eta'}{\sqrt 3}{\bf q}= g_8 {\bf \bar
q}\left(\Phi+\zeta\frac{\eta'}{\sqrt 3}I \right) {\bf q}=g_8 {\bf
\bar q}\left(\Phi'\right) {\bf q}\,, \ee where $\zeta=g_1/g_8$,
$g_1$ and $g_8$ are the coupling constants for the singlet and
octet GBs, respectively, $I$ is the $3\times 3$ identity matrix.
The GB field which includes the octet and the singlet GBs is
written as \bea
 \Phi' = \left( \ba{ccc} \frac{\pi^0}{\sqrt 2}
+\beta\frac{\eta}{\sqrt 6}+\zeta\frac{\eta^{'}}{\sqrt 3} & \pi^+
  & \alpha K^+   \\
\pi^- & -\frac{\pi^0}{\sqrt 2} +\beta \frac{\eta}{\sqrt 6}
+\zeta\frac{\eta^{'}}{\sqrt 3}  &  \alpha K^0  \\
 \alpha K^-  &  \alpha \bar{K}^0  &  -\beta \frac{2\eta}{\sqrt 6}
 +\zeta\frac{\eta^{'}}{\sqrt 3} \ea \right) {\rm and} ~~~~q =\left( \ba{c} u \\ d \\ s \ea
\right)\,. \eea

SU(3) symmetry breaking is introduced by considering $M_s >
M_{u,d}$ as well as by considering the masses of GBs to be
nondegenerate
 $(M_{K,\eta} > M_{\pi})$ {\cite{{song},{johan},{cheng1}}}, whereas
  the axial U(1) breaking is introduced by $M_{\eta^{'}} > M_{K,\eta}$
{\cite{{cheng},{song},{johan},{cheng1}}}. The parameter
$a(=|g_8|^2$) denotes the probability of chiral fluctuation  $u(d)
\rightarrow d(u) + \pi^{+(-)}$, whereas $\alpha^2 a$, $\beta^2 a$
and $\zeta^2 a$ respectively denote the probabilities of
fluctuations $u(d) \rightarrow s + K^{-(0)}$, $u(d,s) \rightarrow
u(d,s) + \eta$,
 and $u(d,s) \rightarrow u(d,s) + \eta^{'}$.

The chiral structure of QCD is known to have intimate connection
with the $\eta$ and $\eta'$ dynamics\cite{etaprob}. Recently, in
the context of $\chi$CQM, Steven D. Bass \cite{bass} has
reiterated in detail the deep relationship of the non-perturbative
aspects of QCD, including gluon anomaly, and the comparatively
large masses of the $\eta$ and $\eta'$ mesons. In particular, he
has emphasized that the gluon degrees of freedom mix with the
flavor singlet Goldstone state to increase the masses of $\eta$
and $\eta'$ through the Witten-Veneziano mass formula
\cite{witten}. Similarly, as shown earlier by Ohta et
al\cite{ohta} and recently advocated by Cheng and Li \cite{cheng}
that $\eta'$ could play an important role in the formulation of
the $\chi$CQM as it may correspond to the non-planar contributions
in the $1/N_c$ expansion. On the other hand, it has recently been
observed on phenomenological grounds \cite{johan} that the new
measurement of both the $\bar u/\bar d$ asymmetry as well as $\bar
u-\bar d$ asymmetry by the NuSea Collaboration \cite{e866} may not
require substantial contribution of $\eta'$. In this context, it
therefore becomes interesting to understand the extent to which
the contribution of $\eta'$ is required in the $\chi$CQM thereby
giving vital clues to the dynamics of non-perturbative regime of
QCD.

The purpose of the present communication is to phenomenologically
estimate the coupling strength of $\eta'$ by carrying out a fine
grained analysis of  ``proton spin problem'' and related issues
within $\chi$CQM$_{{\rm config}}$ by including the latest data. In
particular, it would be interesting to fine tune the contribution
of $\eta'$ by studying its implications on spin polarization
functions,  baryon octet magnetic moments and the quark
distribution functions.

The details of $\chi$CQM$_{{\rm config}}$ have already been
discussed in Ref. \cite{hd}, however to facilitate the discussion
as well as for the sake of readability of the manuscript, some
essential details of $\chi$CQM with configuration mixing have been
presented in the sequel. The most general configuration mixing,
generated by the chromodynamic spin-spin forces
\cite{{dgg},{Isgur},{yaouanc}}, in the case of octet baryons  can
be expressed as \cite{{Isgur},{yaouanc},{full}}
\be
|B \rangle=\left(|56,0^+\rangle_{N=0} \cos \theta +|56,0^+
\rangle_{N=2} \sin \theta \right) \cos \phi +
\left(|70,0^+\rangle_{N=2} \cos \theta^{'} +|70,2^+\rangle_{N=2}
\sin \theta^{'} \right) \sin \phi\,, \label{full mixing} \ee where
$\phi$ represents the $|56\rangle-|70\rangle$ mixing, $\theta$ and
$\theta^{'}$ respectively correspond to the mixing among
$|56,0^+\rangle_{N=0}-|56,0^+ \rangle_{N=2}$ states and
$|70,0^+\rangle_{N=2}-|70,2^+\rangle_{N=2}$ states. For the
present purpose, it is adequate {\cite{{yaouanc},hd,{mgupta1}}} to
consider the mixing only between $|56,0^+ \rangle_{N=0}$ and the
$|70,0^+\rangle_{N=2}$ states and the corresponding ``mixed''
octet of baryons is expressed as
\begin{equation}
|B\rangle \equiv \left|8,{\frac{1}{2}}^+ \right> = \cos \phi
|56,0^+\rangle_{N=0} + \sin \phi|70,0^+\rangle_{N=2}\,,
\label{mixed}
\end{equation}
for details of the  spin, isospin and spatial parts of the
wavefunction,  we  refer the reader to reference
{\cite{{yaoubook}}.

To study the variation of the $\chi$CQM parameters and the role of
the coupling strength of $\eta'$ in obtaining the fit, one needs
to formulate the experimentally measurable quantities having
implications for these parameters as well as dependent on the
unpolarized quark distribution functions and the  spin
polarization functions. We first calculate the spin polarizations
and the related quantities which are affected by the ``mixed''
nucleon. The spin structure of a nucleon is defined as
\cite{{cheng},{song},{johan},hd}
\be
\hat B \equiv \langle B|N|B\rangle, \ee where $|B\rangle$ is the
nucleon wavefunction defined in Eq. (\ref{mixed}) and $N$ is the
number operator given by
\be
 N=n_{u^{+}}u^{+} + n_{u^{-}}u^{-} +
n_{d^{+}}d^{+} + n_{d^{-}}d^{-} + n_{s^{+}}s^{+} +
n_{s^{-}}s^{-}\,, \ee where $n_{q^{\pm}}$ are the number of
$q^{\pm}$ quarks. The spin structure of the ``mixed'' nucleon,
defined through the
 Eq. (\ref{mixed}), is given by
\be
 \left\langle 8,{\frac{1}{2}}^+|N|8,{\frac{1}{2}}^+\right\rangle=\cos^2 \phi
\langle 56,0^+|N|56,0^+\rangle+\sin^2
\phi\langle70,0^+|N|70,0^+\rangle. \label{spinst} \ee The
contribution to the proton spin in $\chi$CQM$_{{\rm config}}$,
given by the spin polarizations defined as $\Delta q=q^+-q^-$, can
be written as

\be
   \Delta u =\cos^2 \phi \left[\frac{4}{3}-\frac{a}{3}
   (7+4 \alpha^2+ \frac{4}{3} \beta^2
   + \frac{8}{3} \zeta^2)\right]+ \sin^2 \phi \left[\frac{2}{3}-\frac{a}{3} (5+2 \alpha^2+
  \frac{2}{3} \beta^2 + \frac{4}{3} \zeta^2)\right], \label {du}\ee
\be
  \Delta d =\cos^2 \phi \left[-\frac{1}{3}-\frac{a}{3} (2-\alpha^2-
  \frac{1}{3}\beta^2- \frac{2}{3} \zeta^2)\right]
+ \sin^2 \phi \left[\frac{1}{3}-\frac{a}{3} (4+\alpha^2+
  \frac{1}{3} \beta^2 + \frac{2}{3} \zeta^2)\right], \label{dd}
  \ee
\be   \Delta s = -a \alpha^2\,. \ee

After having formulated the spin polarizations of various  quarks,
we consider several measured quantities which are expressed in
terms of the above mentioned spin  polarization functions. The
flavor non-singlet components  $\Delta_3$ and $\Delta_8$, usually
calculated in the $\chi$CQM, obtained from the neutron
$\beta-$decay and the weak decays of hyperons are respectively,
\bea \Delta_3= \Delta u-\Delta d =-\frac{1}{9}(5 \cos^2 \phi +
\sin^2 \phi)(-3+ a(3 + 3\alpha^2 + \beta^2 + 2\zeta^2))\,, \\
\Delta_8= \Delta u+\Delta d-2 \Delta s = -\frac{1}{3}(-3+ a(9
-3\alpha^2 + \beta^2 + 2\zeta^2))\,. \label{bsr} \eea The flavor
non-singlet component $\Delta_3$ is related to the well known
Bjorken sum rule \cite{bjor}. Another quantity which is usually
evaluated is the flavor singlet component related to the total
quark spin content as
\be
\Delta \Sigma= \frac{1}{2}(\Delta u+\Delta  d+\Delta
s)=-\frac{1}{6}(-3+ a(9 +6\alpha^2 + \beta^2 + 2\zeta^2))\,, \ee
in the $\Delta s=0$ limit, this reduces to the Ellis-Jaffe sum
rule \cite{ellis}.

Apart from the above mentioned spin polarization we have also
considered the quark distribution functions which have
implications for $\zeta$ as well as for other $\chi$CQM
parameters. For example, the antiquark flavor contents of the
``quark sea'' can be expressed as \cite{{cheng},{song},{johan},hd}
\be \bar u =\frac{1}{12}[(2 \zeta+\beta+1)^2 +20] a\,,~~~~  \bar d
=\frac{1}{12}[(2 \zeta+ \beta -1)^2 +32] a\,,~~~~
 \bar s =\frac{1}{3}[(\zeta -\beta)^2 +9 {\alpha}^{2}] a\,, \ee
 and
 \be
 u-\bar u=2\,,~~~~~d-\bar d=1\,,~~~~~s-\bar s=0\,.
 \ee

The deviation of Gottfried  sum rule \cite{GSR}, related to the
$\bar u(x)$ and $\bar d(x)$ quark distributions, is expressed as
\be
I_G =\frac{1}{3}+\frac{2}{3} \int_0^1 {[\bar u(x)-\bar d(x)]
dx}=0.254\pm 0.005\,. \label{devgsr} \ee In terms of the symmetry
breaking parameters $a$, $\beta$ and $\zeta$, this deviation is
given as
\be
\left[I_G-\frac{1}{3}\right]= \frac{2}{3}\left[\frac{a}{3}( 2
\zeta+ \beta-3)\right]\,. \label{zeta} \ee Similarly, $\bar u/\bar
d$ {\cite{e866,baldit}} measured through  the ratio of  muon pair
production cross sections  $\sigma_{pp}$ and $\sigma_{pn}$, is
expressed in the present case as follows
\be
\bar u/\bar d=\frac{(2 \zeta +\beta +1)^2+20}{(2 \zeta+ \beta-1)^2
+32}\,. \label{u/d}\ee Some of the other important quantities
depending on the quark distribution functions which are usually
discussed in the literature are the fractions of the quark content
and are defined as follows  \be f_q=\frac{q+\bar q}{[\sum_{q}
(q+\bar q)]}\,,~~~~ f_3= f_u-f_d\,, ~~~~f_8= f_u+f_d-2 f_s \,. \ee

Before discussing the results, we first discuss the general
constraints on the $\chi$CQM$_{{\rm config}}$ parameters due to
the spin polarization functions and quark distribution functions.
The $\chi$CQM$_{{\rm config}}$ involves five parameters, $a$,
$\alpha$, $\beta$, $\zeta$ and $\phi$, the mixing angle $\phi$ is
fixed from the consideration of neutron charge radius
\cite{{yaouanc},{full},{neu charge}} as was done in our earlier
calculations \cite{hd}. In the $\chi$CQM, there is a broad
consensus \cite{cheng,song,johan,hd} about the parameters $a$,
$\alpha^2 a$, $\beta^2 a$ and $\zeta^2 a$, representing the
probabilities of fluctuations to pions, $K$, $\eta$ and
$\eta^{'}$, following the hierarchy $a > \alpha^2 a > \beta^2 a >
\zeta^2 a$. The parameter $\zeta$, representing the contribution
of the singlet GB $\eta'$ in the $\chi$CQM, as well as  the
parameters $a$, $\alpha$ and $\beta$ cannot be fixed independently
from the spin polarization functions and quark distribution
functions. To begin with, we have carried out a broader analysis
wherein we have attempted to find the ranges of the $\chi$CQM
parameters from $\Delta u$, $\Delta_3$, $\bar u-\bar d$, $\bar
u/\bar d$.

The range of the coupling breaking parameter $a$ can be easily
found by considering the expression of the spin polarization
function $\Delta u$ (Eq. (\ref{du})), by giving the full variation
of parameters $\alpha$, $\beta$ and $\zeta$ from which one finds
$0.10 \lesssim a \lesssim 0.14$, in agreement with the values
considered in other similar calculations
\cite{cheng,song,johan,hd}. The range of the parameter $\zeta$ can
be found from the expression of $\bar u/\bar d$ (Eq. (\ref{u/d}))
which involves only $\beta$ and $\zeta$. Using the possible range
of $\beta$, i.e. $0<\beta<1$ as well as the latest experimental
measurement  of $\bar u/\bar d$ \cite{e866}, one finds $-0.70
\lesssim \zeta \lesssim -0.10$. The parameter $\zeta$, except in
the quark distribution functions, occurs as $\zeta^2$ in the spin
polarization functions and the calculations of spin dependent
quantities, therefore, for the purpose of discussion, the
constraint on $\zeta$ can be expressed as $0.10 \lesssim |\zeta|
\lesssim 0.70$. The range of $\beta$ can be found by using the
$\bar u-\bar d$ asymmetry representing the violation of Gottfried
sum rule \cite{GSR}. In terms of the $\chi$CQM parameters, the
$\bar u-\bar d$ asymmetry can be expressed as \be \bar u-\bar
d=\frac{a}{3}(2 \zeta+\beta-3)\,. \label{asym} \ee Using the above
found ranges of $a$ and $\zeta$ as well as the latest measurement
of $\bar u-\bar d$ asymmetry \cite{e866}, $\beta$ comes out to be
in the range $0.2\lesssim \beta \lesssim 0.7$. Similarly, the
range of $\alpha$ can be found by considering the flavor
non-singlet component $\Delta_3$ (=$\Delta u-\Delta d$) and it
comes out to be $0.2 \lesssim \alpha \lesssim 0.5$.

Our gross analysis as well as earlier analyses
\cite{cheng,song,johan,hd} brings out that $a$, representing the
``pion'' fluctuation, expectedly is the most important parameter.
The parameters $\alpha$ and $\beta$ are nearly equal whereas the
parameter $\zeta$ is strongly coupled to $a$ through the $\bar
u-\bar d$ asymmetry,. After finding the ranges of the coupling
breaking parameters, we have carried out a fine grained analysis,
first by fitting $\Delta u$, $\Delta d $ and $\Delta_3$ \cite{PDG}
and then calculating the other spin polarization functions, quark
distribution functions and baryon octet magnetic moments with the
same values of parameters. The analysis consist of several fits
which have been carried out to constrain the coupling strength of
$\eta'$.

In several previous calculations the parameter $a$ has usually
been fixed to be 0.1 \cite{cheng,song,johan,hd}, therefore we have
also attempted to fit the data by keeping $a=0.1$ and treating the
other three parameters to be free leading to $|\zeta|=0.65$,
$\alpha=0.4$ and $\beta=0.7$. A similar analysis has also been
carried out for $|\zeta|=0.70$, corresponding to the extreme value
of $\zeta$, whereas  $\alpha$ and $\beta$ are treated as free
parameters, yielding $\alpha=0.4$ and $\beta=0.6$. The parameter
$a$, as noted earlier also \cite{cheng,song,johan,hd}, plays a
very important role in carrying out the overall fit in the
$\chi$CQM, therefore we have attempted a fit by also keeping the
parameter $a$ to be free. However, to keep the escalation of the
parameters under check, we have carried out a fit wherein $a$,
$\alpha$ and $\zeta$ are treated as free parameters whereas
$\beta$ is taken to be equal to $\alpha$, in accordance with our
broader analysis as well as suggested by Cheng and Li
\cite{cheng}, leading to $a=0.13$, $|\zeta|=0.10$,
$\alpha=\beta=0.45$. Interestingly, this fit turns out to be as
good as the fit wherein all the four parameters are treated free,
henceforth, this fit would be referred to as the best fit. To have
a deeper understanding of the $\zeta$ values corresponding to the
best fit, we have also carried out our analysis where there is no
contribution of the singlet GB ($\zeta=0$) with $a$, $\alpha$ and
$\beta$ being treated free, yielding the values $a=0.14$,
$\alpha=0.4$ and $\beta=0.2$. For the purpose of discussion, even
for the quark distribution functions, the $\zeta$ values
corresponding to $|\zeta|=0.65$ and $|\zeta|=0.70$ would be
referred to as higher values of $\zeta$ (Cases I and II
respectively) whereas $\zeta=0$ and $|\zeta|=0.10$ would be
referred to as lower values of $|\zeta|$. The analysis with
$\zeta=0$ is referred to as Case III whereas our best fit, for the
sake of uniformity, is referred to as Case IV.

In Table \ref{spin}, we have presented the results of our fits
mentioned above. A cursory look at the table immediately brings
out that our best fit clearly has much better overlap with the
data compared to all other fits presented in the table. However,
before getting into the detailed comparison of this fit with the
data, we compare the two fits corresponding to the higher values
of $|\zeta|$ with each other, for example $|\zeta|=0.65$ and
$|\zeta|=0.70$, primarily for the purpose of understanding the
role of $\eta'$. The two cases expectedly are not very different
from each other, however on closer examination one finds that Case
I is uniformly better than II even though by a very small amount.
The slight improvement observed in the cases of $\Delta u$,
$\Delta d$ and $\Delta_8$ can be understood as a consequence of
the change in the value of the parameter $\zeta$ as the above
mentioned quantities can be seen to have weak dependence on
$\alpha$ and $\beta$ in comparison to $\zeta$. Therefore, we tend
to conclude that lowering of $|\zeta|$ leads to better overlap
with data. This becomes all the more clear when one notes that in
case of our best fit, the $|\zeta|$ value is 0.10, much lower
compared to $|\zeta|=0.65$ and $|\zeta|=0.70$ . This is further
borne out by the fact that keeping $a=0.13$ and $\alpha=\beta=
0.45$, however changing $|\zeta|$ towards higher values, worsens
the fit compared to our best fit, nevertheless the fit remains
better compared with higher values of $|\zeta|$.

In the case of our best fit, we are able to obtain an excellent
fit for $\Delta u$, $\Delta d$ and $\Delta_8$, in contrast with
the other cases considered here. It needs to be mentioned that
$\Delta_8$ cannot be fitted for  "higher values" of $|\zeta|$ even
after scanning the entire parameter space for $a$, $\alpha$ and
$\beta$, suggesting that only the lower values of $|\zeta|$ are
compatible with data. This can be easily understood when one
closely examines the expression for $\Delta_8$ wherein one finds
that it is completely dominated by the parameter $a$. As already
observed, the parameters $a$ and $\zeta$ are related through $\bar
u-\bar d$ asymmetry, therefore while fitting $\Delta_8$, requiring
relatively higher values of $a$, leads to only lower values of
$|\zeta|$. One may wonder why $\Delta_3(=\Delta u-\Delta d$) gets
fitted for all the fits. This can be easily understood when one
realizes that even though $\Delta u$ and $\Delta d$ have
significant dependence on $a$ but this dependence gets cancelled
in $\Delta_3$. Similarly, one can understand that $\Delta_3$ also
does not have any significant dependence on $\zeta$ because of the
above mentioned cancellation.

After finding that the lower values of $|\zeta|$ giving much
better overlap with data, one would like to check whether the same
remain true in the case of octet magnetic moments or not. In this
case also we find that our best fit values corresponding to lower
value of $|\zeta|$ (Case IV) has a much better overlap with data
compared to the higher values of $|\zeta|$. Specifically, in the
case of $\mu_{n}$, $\mu_{\Sigma^-}$, $\mu_{\Sigma^+}$ and
$\mu_{\Xi^-}$, the fit with lower value of $|\zeta|$ scores over
higher values of $|\zeta|$ in a marked manner. Again this remains
true when we scan the entire parameter space. It may also be
mentioned that $\mu_{\Xi^-}$ is a difficult case in most of the
quark models \cite{qms}.

To understand the extent to which the value of $|\zeta|$ can be
lowered, we now consider Case III which corresponds to the singlet
GB being absent ($\zeta=0$). A general look at the table indicates
that in the absence of the singlet GB, even though the other
parameters being kept free, we find that the data clearly
indicates preference for a small non-zero value of $|\zeta|$. For
example, in the case $\mu_{p}$, $\mu_{n}$, $\mu_{\Sigma^-}$,
$\mu_{\Sigma^+}$, $\mu_{\Xi^o}$ and $\mu_{\Xi^-}$, the
$|\zeta|=0.10$ fit has clearly better overlap with data as
compared to $\zeta=0$. In the case of $\mu_{\Xi^-}$, one may
wonder, why such a small change in the value of $\zeta$ affects
the fit in a marked manner. The significant improvement in the
case of $\mu_{\Xi^-}$ for the best fit, compared to the $\zeta=0$
case, can be easily understood when one realizes that the magnetic
moments have been formulated by considering the valence, sea and
orbital contributions with appropriate signs \cite{hd}. A small
change in the value of $\zeta$ changes the sea and orbital
contributions which constructively add to the total contribution,
hence justifying the better agreement with data in the case of
$|\zeta|=0.10$ compared to $\zeta=0$. In a similar manner one can
understand the better agreement achieved here with data for the
case of other magnetic moments.

 In Table \ref{quark}, we have presented the
results corresponding to quark distribution functions for
different fits. In this case also the fit for the lower values of
$|\zeta|$ is better as compared to the higher values in the case
of $\bar u/\bar d$. It is interesting to observe that even for a
small deviation in the value of $\zeta$, $f_s$ gets affected
significantly, therefore a measurement of $f_s$ would give a very
strong signal about the coupling strength of $\eta'$ in the
$\chi$CQM. It may be mentioned that our conclusion regarding the
small but non-zero value of $|\zeta|$ being preferred over
$\zeta=0$  is not only in agreement with latest $\bar u-\bar d$
measurement \cite{e866} but is also in agreement with the
conclusions of Ohlsson {\it et. al.} \cite{johan}.

The conclusions arrived at above can perhaps be understood from a
more theoretical point of view also. In the $\chi$CQM, it is
difficult to think of a mechanism wherein the contribution of
$\eta'$ or the singlet GB becomes zero, whereas a small value of
the coupling strength of $\eta'$ from phenomenological
considerations is in agreement with the arguments advanced by S.
Bass \cite{bass} and emphasized in the beginning. This is also in
agreement with the suggestion of  Cheng and Li \cite{cheng}
wherein they have suggested that the coupling strength of the GB
corresponding to the pion, $K$, $\eta$ and $\eta'$ mesons are
inversely proportional to the square of their respective masses
and are respectively of the order $a \alpha^2 \sim 0.02$, $a
\beta^2 \sim 0.02$ and $a \zeta^2 \sim 0.001$ for $a \sim 0.13$.
This strangely agrees with the values obtained through our best
fit as well as suggests the equality of the coupling strength of
$\alpha$ and $\beta$.

In the Table \ref{spin}, we have also presented the results of our
calculations for the flavor singlet component of the spin of
proton $\Delta \Sigma$, however we have not discussed its
implications for different cases. It has already been discussed
earlier in $\chi$CQM \cite{hdgluon} that $\Delta \Sigma$ receives
contributions from various sources such as valence quarks, quark
sea, gluon polarization etc.. In the context of $\chi$CQM, it
seems that gluon anomaly  not only contributes to the gluon
polarization but also is responsible for the large mass of $\eta'$
\cite{bass}. As a result, its contribution to $\Delta \Sigma$ gets
correspondingly reduced compared to the contributions of valence
quarks and quark sea. In this context, we would like to mention
that the above non-zero but small contribution of $\eta'$ looks to
be well in agreement with our earlier calculations regarding the
partitioning of the nucleon spin \cite{hdgluon}.

To summarize,  with a view to  phenomenologically estimating the
coupling strength of the singlet Goldstone boson $\eta'$ in
$\chi$CQM$_{{\rm config}}$,  we have carried out a detailed
analysis  using the latest data regarding $\bar u-\bar d$
asymmetry, the spin polarization functions and the baryon octet
magnetic moments. The $\chi$CQM$_{{\rm config}}$ involves the
parameters $a$, $\alpha^2 a$, $\beta^2 a$ and $\zeta^2 a$,
representing respectively the probabilities of fluctuations to
pions, $K$, $\eta$ and $\eta^{'}$, as well as $\phi$ which is
fixed from the consideration of neutron charge radius
\cite{{yaouanc},{full},{neu charge}}. As a first step of the
analysis, we have found from broad considerations the required
ranges of these parameters using the data pertaining to $\Delta
u$, $\Delta_3$, $\bar u-\bar d$, $\bar u/\bar d$ etc.. The ranges
obtained are $0.10\lesssim a \lesssim 0.14$, $0.2\lesssim \alpha
\lesssim 0.5$,  $0.2\lesssim \beta \lesssim 0.7$ and $0.10\lesssim
|\zeta| \lesssim 0.70$. After obtaining the ranges, analysis has
been carried out corresponding to four different sets of the
$\chi$CQM parameters within the ranges mentioned above as well as
keeping in mind the earlier analyses done in this regard. In the
first case, the pion fluctuation parameter $a$ is taken as 0.1,
the value considered by several earlier analyses
\cite{cheng,song,johan,hd}, whereas $\Delta u$, $\Delta_3$, $\bar
u-\bar d$, $\bar u/\bar d$ etc. are fitted by treating the other
three parameters to be free. This analysis yields $|\zeta|=0.65$,
$\alpha=0.4$ and $\beta=0.7$ and is referred to as Case I. A
similar analysis has also been also been carried out by taking
$a=0.1$ and $|\zeta|=0.70$ whereas the parameters $\alpha$ and
$\beta$ are treated free, yielding $\alpha=0.4$ and $\beta=0.6$
and is referred to as Case II. Our best fit is obtained by varying
$a$, $\zeta$ and $\alpha$, the parameter $\beta$ is taken to be
equal to $\alpha$, in accordance with the suggestions of Cheng and
Li \cite{cheng}. Interestingly, this fit is not affected when
$\alpha$ and $\beta$ are not taken equal and the best fit values
of the parameters are $a=0.13$, $|\zeta|=0.10$,
$\alpha=\beta=0.45$. We have also carried out a fit where there is
no contribution of the singlet GB ($\zeta=0$) and $a$, $\alpha$ as
well as $\beta$ are treated free, yielding $a=0.14$, $\alpha=0.4$
and $\beta=0.2$. The analysis with $\zeta=0$ is referred to as
Case III whereas the best fit is referred to as Case IV. The
values of the parameter corresponding to $|\zeta|=0.65$ and
$|\zeta|=0.70$ are referred to as higher values of $|\zeta|$
whereas those corresponding to $\zeta=0$ and $|\zeta|=0.10$ are
referred to as lower values of $|\zeta|$. A comparison of all the
fits clearly shows that our best fit is not only better than other
fits carried out here but also provides an excellent overall fit
to the data particularly in the case of $\Delta u$, $\Delta d$,
$\Delta_8$, $\mu_{n}$, $\mu_{\Sigma^-}$, $\mu_{\Sigma^+}$,
$\mu_{\Xi^o}$ and $\mu_{\Xi^-}$.

As already discussed in detail, comparison of Cases I and II
suggest that in general lower values of $|\zeta|$ are preferred
over higher values of $|\zeta|$. This is borne out clearly by a
closer look at our best fit. In fact, it is interesting to mention
that any change of $|\zeta|$ from the value $0.10$, worsens the
fit. This can further be understood very easily by analyzing $\bar
u-\bar d$ asymmetry, $\Delta_8$ as well as $\mu_{n}$,
$\mu_{\Sigma^-}$ and $\mu_{\Xi^-}$. The analysis of $\bar u-\bar
d$ asymmetry clearly suggests that the values of $a$ and $\zeta$
are strongly coupled to each other independent of the values of
$\alpha$ and $\beta$. Further, the parameter $a$ plays the most
important role in fitting the polarization functions $\Delta u$,
$\Delta d$, $\Delta_3$ and $\Delta_8$ whereas the role of $\alpha$
and $\beta$ is much less significant. Thus, one finds that the
higher values of $a$ have to go with the lower values of $|\zeta|$
and vice-versa. Unlike the Case I, in the best fit we are able to
fit $\Delta u$, $\Delta d$, $\Delta_3$ and $\Delta_8$
simultaneously. Similar conclusion can be arrived at from an
analysis of the magnetic moments $\mu_{n}$, $\mu_{\Sigma^-}$,
$\mu_{\Sigma^+}$ and $\mu_{\Xi^-}$. From the above discussion, one
can conclude that the data very strongly suggests that lower
values of $|\zeta|$ and correspondingly higher values of $a$ are
preferred over the higher values of $|\zeta|$ and lower values of
$a$.

To further understand the coupling strength of $\eta'$ in
$\chi$CQM, we have carried out in Case III an analysis where the
contribution of $\eta'$ is taken to be zero. Interestingly, we
find that the fit obtained in this case cannot match our best fit
even if the other parameters are treated completely free,
suggesting that the $\zeta=0$ case can be is excluded
phenomenologically. This conclusion regarding the exclusion of the
singlet GB also looks to be in agreement with the theoretical
considerations based on the arguments of Cheng and Li \cite{cheng}
and those of S. Bass \cite{bass}. It seems that the
phenomenological analyses of spin polarization functions, quark
distribution functions and baryon octet magnetic moments, strongly
suggest a small but non-zero value of $|\zeta|$ within the
dynamics of chiral constituent quark model, suggesting an
important role for $\eta'$ in the non-perturbative regime of QCD.
This fact perhaps can be substantiated by a measurement of the
quark distribution function $f_s$ which shows a strong dependence
on the value of $\zeta$.

 \vskip .2cm
 {\bf ACKNOWLEDGMENTS}\\
H.D. would like to thank DST (OYS Scheme), Government of India,
for financial support and the chairman, Department of Physics, for
providing facilities to work in the department.

\pagebreak

\begin{table}
\begin{center}
\begin{tabular}{cccccc}
Parameter & Data  & \multicolumn{4}{c}{$\chi$CQM$_{{\rm config}}$}
\\  \cline{3-6}
& & Case I& Case II & Case III & Case IV \\ \cline{3-6} & &
$|\zeta|=0.65$ & $|\zeta|=0.70$ & $\zeta=0$ & $|\zeta|=0.10$\\
 &     & $a=0.1$ &
$a=0.1$ & $a=0.14$ & $a=0.13$\\ & & $\alpha=0.4$ & $\alpha=0.4$ &
$\alpha=0.4$ & $\alpha=0.45$ \\ & & $\beta=0.7$ & $\beta=0.6$ &
$\beta=0.2$ & $\beta=0.45$
\\
 \hline

$\Delta u$ & 0.85 $\pm$ 0.05 \cite{emc} & 0.947 & 0.955 & 0.925 &
0.913
\\ $\Delta d$ & $-$0.41  $\pm$ 0.05 \cite{emc}   & $-$0.318 &
$-0.312$& $-$0.352& $-$0.364 \\ $\Delta s$ &$-$0.07  $\pm$ 0.05
\cite{emc} &$-$0.02&$-$0.02 &$-$0.02 &$-$0.02 \\ $\Delta_3$ &
1.267 $\pm$ 0.0035 \cite{PDG} & 1.267 & 1.267 & 1.267 & 1.267\\
$\Delta_8$ & 0.58  $\pm$ 0.025 {\cite{PDG}} &0.67 & 0.68& 0.61 &
0.59 \\ $\Delta \Sigma$ & 0.19 $\pm$ 0.025 {\cite{PDG}}& 0.31&
0.31 &0.28 & 0.27
\\ \hline $\mu_{p}$ & 2.79$\pm$0.00 {\cite{PDG}} & 2.80 & 2.80 & 2.83 & 2.81 \\
 $\mu_{n}$ & $-1.91\pm$0.00 {\cite{PDG}}&  $-$1.99 & $-$2.00 &
$-$2.16 & $-$1.96  \\ $\mu_{\Sigma^-}$ & $-1.16\pm$0.025
{\cite{PDG}}& $-$1.20 & $-$1.21& $-$1.32& $-$1.19 \\
$\mu_{\Sigma^+}$ & 2.45$\pm$0.01 {\cite{PDG}} & 2.43& 2.42& 2.53 &
2.46
\\ $\mu_{\Xi^o}$ & $-1.25\pm$0.014 {\cite{PDG}} &  $-$1.24 & $-$1.23 & $-$1.33 & $-$1.26 \\
$\mu_{\Xi^-}$ & $-0.65\pm$0.002 {\cite{PDG}} & $-$0.56 &$-$0.57&
$-$0.67 & $-$0.64
\\
\end{tabular}
\end{center}
 \caption{The calculated values of the spin polarization
functions and  baryon octet  magnetic moments for different cases.
The value of the mixing angle $\phi$ is taken to be $20^o$. }
\label{spin}
\end{table}

\begin{table}
\begin{center}
\begin{tabular}{cccccc}

Parameter & Data  & \multicolumn{4}{c}{$\chi$CQM} \\ \cline{3-6} &
& Case I& Case II & Case III & Case IV \\ \cline{3-6} & &
$|\zeta|=0.65$ & $|\zeta|=0.70$ & $\zeta=0$ & $|\zeta|=0.10$\\ & &
$a=0.1$ & $a=0.1$ & $a=0.14$& $a=0.13$ \\ & & $\alpha=0.4$ &
$\alpha=0.4$ & $\alpha=0.4$ & $\alpha=0.45$ \\ & & $\beta=0.7$ &
$\beta=0.6$ & $\beta=0.2$ & $\beta=0.45$
\\ \hline $\bar u$ & $-$ & 0.168 & 0.167&0.250 & 0.233 \\

$\bar d$ & $-$ &  0.288 & 0.293&  0.366& 0.350 \\

$\bar s$ & $-$   &   0.108 & 0.104& 0.07& $0.07$ \\

$\bar u-\bar d$ & $-0.118 \pm$ 0.015 \cite{e866} & $-0.120$&
$-0.127$ & $-0.116$ & $-0.117$ \\

$\bar u/\bar d$ & 0.67 $\pm$ 0.06 {\cite{e866}}  & 0.58 &0.57
&0.68& 0.67  \\

$I_G$ & 0.254  $\pm$ 0.005  & 0.253 &0.248&0.255 & 0.255  \\

$f_u$ &$-$   &   0.655 & 0.654 &0.677& 0.675 \\

$f_d$ &$-$ &  0.442 &0.445& 0.470 & 0.466 \\

$f_s$ &  0.10 $\pm$ 0.06 {\cite{ao}}  &  0.061 & 0.058 &0.039 &
0.039
\\

$f_3$  &$-$ & 0.213 & 0.209 & 0.207 & 0.209\\

$f_8$  &$-$ & 0.975 &0.982&  1.07& 1.06 \\

$f_3/f_8$ & 0.21 $\pm$ 0.05 {\cite{cheng}}  &  0.22 & 0.21 & 0.19
& 0.20
\\

\end{tabular}
\end{center}
\caption{The calculated values of the quark flavor distribution
functions for different cases.} \label{quark}
\end{table}

\end{document}